\documentstyle[11pt,sokeriaur,twoside]{article}
\def\edcomment#1{\iffalse\marginpar{\raggedright\sl#1\/}\else\relax\fi}
\reversemarginpar
\begin{document}
\title{Planetary Nebulae in the Scheme of Binary Evolution}
\author{Noam Soker}
\affil{Dept. of Physics, University of Haifa-Oranim, Tivon 36006, Israel}

\begin{abstract}

In this review I present the binary model for the shaping
of planetary nebulae (PNe) as I view it, in the context of
historical evolution of other models for the shaping of PNe over 
more than 30 years. 
In describing the binary model, I concentrate on 
works published since the last IAU meeting on PNe. 
 I think stellar companions are behind the shaping of bipolar PNe,
i.e., having two lobes with an equatorial waist between them,
and extreme elliptical PNe, e.g., having a dense equatorial ring,
but no lobes. 
The question of whether a planet is required to spin the progenitor
in order to form a moderate elliptical PN, or whether a single
stars can form moderate elliptical PNe, I consider to be
an open question. 

\end{abstract}

\section{Introduction}

The rich variety of planetary nebula (PN) shapes have attracted attention
for a long time. 
Different mechanisms were discussed to form the axisymmetrical
structures of PNe.
Magnetic fields, via different mechanisms, were, and continue to be, 
quite popular.
The idea that the galactic magnetic field shapes PNe was raised
several times (e.g., Grinin \& Zvereva 1968), but was always
`killed', recently by Corradi, Aznar, \& Mampaso (1998).
Internal magnetic fields, via several different mechanisms, have been 
discussed for more than 40 years  (e.g., Gurzadyan 1962; Woyk 1968 
[note his drawings!]; Pascoli 1985, 1997; Chevalier \& Luo 1994; 
Garc\'{\i}a-Segura 1997; Garc\'{\i}a-Segura et al.\ 1999; 
Garc\'{\i}a-Segura, \& L\'opez 2000; Garc\'{\i}a-Segura, L\'opez, 
\& Franco 2001; 
Matt et al.\ 2000; Blackman et al.\ 2001; Gardiner, \& Frank 2001; 
see also review by Garc\'{\i}a-Segura in these proceedings). 
These models assume that the magnetic field comes from the AGB progenitor,
and attribute dynamical effects to the magnetic field, i.e.,
the magnetic pressure and/or tension become comparable to that of
the thermal pressure.
 Some magnetic effects were criticized during these years 
(e.g., Menzel 1968; Soker \& Zoabi 2002 and references therein). 
In other models, the magnetic field has a secondary role. 
In several papers (e.g., Soker 1998b, 2000, 2001b ; Soker, \& Clayton 1999; 
Soker \& Zoabi 2002) I propose that  the large-scale stellar magnetic 
field is strong enough for the formation of magnetic cool spots on the 
AGB stellar surface.
The spots may regulate dust formation, hence mass loss rate,
leading to axisymmetric mass loss and the formation of elliptical PNe.
Despite its role in forming cool spots, the large scale
magnetic field is too weak to play a dynamic role and directly
influence the wind from the AGB star, as required by the dynamic models.

A popular mechanism for shaping PNe is the so called interacting
winds model (see review by Frank 1999). 
The basic model assumes mass concentration in the equatorial plane
from the AGB phase. Later, the fast wind from the central star of
the PN accelerates the nebula, with higher velocities and distances
attended in the polar directions, forming elliptical or bipolar
PNe. 
 Ionization plays a significant role in the evolution
(Mellema 1995; Mellema \& Frank 1995). 
The idea of mass concentration in the equatorial plane is old;
Khromov, \& Kohoutek (1968) suggested that mass concentration in the
equatorial plane can focus the ionization radiation to the poles. 
However, it became a popular idea with the seminal work of
Balick (1987).
Interaction of the fast and slow winds definitely occurs in PNe,
and is successful in explaining some properties of PNe.
However, it cannot explain all properties  
in many of the observed bipolar PNe, e.g., the high momentum
and kinetic energy observed in several bipolar PNe and proto-PNe
(Bujarrabal et al.\ 2001).
Other problems are mention by Frank (2000; also Balick 2000).  

Other, less popular, mechanisms were proposed in the literature, e.g.,
a circumstellar disk that survived from the pre-main sequence
phase (e.g., Kastner et al.\ 1996).

My view (Soker \& Harpaz 1992; Soker 1998b;  Soker \& Zoabi 2002) 
is that a stellar binary companion is required to spin up
the AGB star in order for the later to have a strong magnetic field.
The equatorial mass concentration required for the interacting wind model
to form bipolar and extreme elliptical PNe 
also seems to require a stellar binary companion (e.g., 
Soker \& Livio 1989; Iben \& Livio 1993),
e.g., as obtained in the simulations of Mastrodemos \& Morris (1998, 1999). 
The companion, then, may cause other effects, e.g., accreting
mass, forming an accretion disk (Mastrodemos \& Morris 1999),
and blowing jets (Morris 1987; Soker \& Rappaport 2000). 

Before further exploring binary models, I clarify some terms.

\section{Claryfing Some Terms}

{\bf Single stars.} Some mechanisms may work for single stars, e.g., 
fast rotation (Garc\'{\i}a-Segura et al.\ 1999; but not their claim for 
critical slow rotation; see Glatzel 1998). 
However, in these cases, I argue (Soker \& Harpaz 1999; Soker 2001b ) 
the envelope posses more angular momentum than a single AGB star can 
maintain from its main sequence phase. Hence the envelope
must be spun-up by a stellar companion. I would refer to these cases
as binary evolution, rather than single star evolution, although the
process itself involves only one star.
\newline
{\bf Magnetic field.} There are some indications for magnetic fields
around AGB stars and in PNe, mainly from maser emission
(e.g.,  Miranda et al.\ 2001).
This does not necessarily support models with dynamical effects of
magnetic fields in the AGB progenitors.
There can be other sources for the magnetic field.
(1) The magnetic field can be localized in specific regions. This is predicted
if there is  a higher mass loss rate from active regions, e.g.,
higher mass loss rate from cool spots (e.g., Soker \& Zoabi 2002).
(2) An accreting companion, whether a WD or a main sequence star, will
amplify the magnetic field. Hence the strong field may result
from an accreting companion.
(3) The magnetic field may result from a magnetically active main
sequence companion.
 Accreting main sequence companions will be spun-up, and may become
magnetically active (Jeffries \& Stevens 1996; Soker \& Kastner 2002).
\newline
{\bf Morphology.}
 I will use the following 4 main axisymmetrical morphology classes.
(i) Bipolar PNe: PNe having two lobes with
and equatorial waist between them. 
(ii) Extreme elliptical PNe: PNe having a large equatorial concentration of 
mass, e.g., a ring, but no, or only small, lobes.
They may possess jets. 
(iii) Moderate elliptical PNe: PNe having a large scale elliptical shape,
with a shallow density variation from equator to poles.
They may possess jets. 
(iv) Spherical PNe: PNe whose entire structure show a general circular
shape. If there are even small parts which are axisymmetric rather than 
spherical,  the PN is not spherical (I will not discuss these any more
here; for some properties of spherical PNe see Soker [2002a ]). 

\section{Binary systems and PN Structures}

The idea that binary systems are behind the axisymmetrical structure of PNe
is old, and was mentioned many times in the literature, more times
than can be mentioned here; a few examples are Fabian \& Hansen (1979), 
Phillips, \& Reay (1983), Miranda (1995),  Kolesnik \& Pilyugin (1986),
and Pollacco \& Bell (1997).  
In particular, Morris (1981, 1987, 1990; Mastrodemos \& Morris 1999) 
and Livio (1982, 1993, 1995, 1997, 2000; Livio, Salzman, \& Shaviv 1979; 
Iben \& Livio 1993; Livio \& Pringle 1996)
were pushing the idea of binary interaction for a long
time, proposing many ideas and models.
In my research on the connection between binary systems and PNe,
I benefited from their works, as well as from  observational works 
by Corradi and Schwarz (e.g., Corradi 1995; Corradi \& Schwarz 1995; 
Corradi et al.\ 1999, 2001; Schwarz et al.\ 1997), and Bond
(e.g., Bond 2000). 
In their works, Corradi and Schwarz nailed down the 
similarity between symbiotic nebulae, which known to be formed by 
binary systems, and bipolar PNe (for other, some  earlier, papers on 
this connection and binarity see Mammano \& Ciatti 1975; 
Cohen et al.\ 1985; Morris 1990; Goodrich 1991; Lee \& Park 1999). 

Main supporting observations for the shaping of some PNe by
binary systems is in Soker (1997; as well as different processes
caused by a companion), and supporting observations for
binary shaping of bipolar PNe are in Soker (1998a; see table 1 there), 
and references in these papers.
I only list several of them here, in addition to new results
(for the explanation of massive progenitor of bipolar PNe see $\S 5$).
(i) 16 PNe with central binary systems, one of them bipolar and
the rest extreme elliptical PNe, are known (Bond 2000). 
(ii) Other PNe and proto-PNe show strong indication for binarity
 (e.g., Rodr\'{\i}guez, Corradi, \& Mampaso 2001;
 the 120 years side to side variation in M2-9,  Doyle et al.\ 2000).
(iii) The similar morphology of bipolar PNe and many symbiotic nebulae.
(iv) Expansion velocities in many bipolar PNe 
proto-PNe are much higher than the escape velocity from AGB stars.
(v) Statistically, binary systems which avoid common envelope
  can account for bipolar PNe (Soker \& Rappaport 2000), as models
  requires (next section). 
  Common envelope systems can account for a large fraction
  of elliptical PNe (Yungelson, Tutukov, \& Livio 1993;
  Han, Podsiadlowski, \& Eggleton 1995).      
However, it seems that binary stellar systems can't account for all
non spherical PNe; either single stars can form moderate elliptical,
or planets are required. 
 
\section{Current Status of Binary Models}

I summarize now my view on the role plaid by stellar companions,
as appeared in several of my papers from the last 5 years.
As is well known, accreting companions can blow jets (Livio 2000),
or if not well collimated, termed collimated fast wind (CFW; Soker
\& Rappaport 2000).
The CFW (or jets) can shape the nebula in several ways.
If the CFW is strong, it will form two lobes (Morris 1987), and
form bipolar PNe. 
A close companion can enhance the equatorial density by spining-up
the AGB star (Soker \& Rappaport 2000) and/or focusing 
the AGB wind (Mastrodemos \& Morris 1999).
Like in symbiotic nebulae, the companion should be outside the 
AGB envelope, at least during part of the evolution.
I therefore argue (Soker 1998a), that most bipolar PNe are formed 
from companions outside the envelope; in a few cases (e.g., NGC 2346) 
the companion can enter the envelope at a late stage.
The different routes that can lead to the rich variety of bipolar
shapes is summarized in Soker (2002b ).
Stellar companions at large orbital separations (tens of AU; see Soker
2001a ) may blow weak CFW, which will form extreme elliptical
PNe, possibly with small lobes. Most extreme elliptical PNe are formed
via common envelope evolution (e.g., Bond \& Livio 1990). 

Binary systems can lead to other effects.
Examples are displacement from axisymmetry (Soker \& Rappaport 2001;
Soker \& Hadar 2002), formation of a circumbinary disk 
(e.g., Van Winckel 1999; Jura, Chen \& Plavchan 2002), and 
leading to backflowing material in the post-AGB phase (Soker 2001c ). 

It is not clear, though, whether binary systems are behind the
multiple-rings found by HST in several PNe and proto-PNe
(e.g., Hrivnak, Kwok, \& Su 2001), directly, by their gravity
and orbital motion, or indirectly, by spining-up the AGB star
(for different views, see Soker 2002c  and Y. Simis in these proceedings).

Because stellar binary companions can't account for all non-spherical
PNe, either single stars can form moderate elliptical PNe,
or, as I tend to think, planets are required to spin-up these AGB stars
(Soker 2001b ). 
Whereas I am confident that binary companions shape bipolar and
extreme elliptical PNe, whether a companion is required to shape
moderate elliptical PNe, I consider an open question.

\section{Massive Progenitors of Bipolar PNe}
 
Although more than three years have past since I published my explanation 
for the observations that bipolar PNe are formed from more massive stars,
annoyingly, this correlation is still, wrongly, used as contradictory to 
binary models.
I presented the explanation of Soker (1998a; described below, but see
Soker 1998a for detail) in my talk; despite that I got 2 questions 
after the talk regarding that same point.
Because of that, I devote a section just for this observation. 

As noted by many surveys (e.g., Zuckerman \& Gatley 1988), bipolar PNe are 
concentrated toward the Galactic plane. This and composition differences 
from elliptical PNe strongly suggest that progenitors of bipolar PNe are 
more massive than those of the other PNe (e.g., Greig 1972;
Acker 1980; Kaler 1983; Torres-Peimbert \& Peimbert 1997). 
As explained in Soker (1998a) the main differences between 
massive ($M \ga 2 M_\odot$ on the main sequence) and low-mass progenitors
are that the radii of the massive progenitors become much larger on the AGB 
than on the RGB, and their envelope mass is much larger than that of 
low-mass progenitors.
Because of their large radius on the RGB, low mass stars interact with 
a companion, that could have form bipolar PN, already on their RGB track. 
Either they lose their envelope and never form a PN, because of their 
low mass envelope, or the companion spirals-in to collide with their 
core, after which the star evolves as rapidly rotating single star, 
hence forming an elliptical PNe.  

Another, more simple, explanation is that more massive stars are more
likely than low mass stars to harbor close massive companions that 
lead to the formation of bipolar PN.
This was suggested for binary systems that go through a stable Roche lobe
overflow (Soker \& Livio 1994). 

I end with a comment I made during a debate conducted at the
Asymmetrical Planetary Nebulae II (APN II) meeting: 
There is a class of systems called massive X-ray binaries, 
with a neutron star and a massive companion.
So all of them contain massive stars;  does this mean that these binary
systems are `single stars'?

\section{Future Works and Predictions}

{\bf Theory. } In order to reproduce bipolar morphologies, there is a need 
to perform 3D gasdynamical simulations where both stars blow wind 
simultaneously: The AGB blows a slow wind, either spherically symmetric
or concentrated to the equatorial plane, while the compact companion,
a WD or a main sequence star, blows a CFW (collimated fast wind, or jets). 
Since the binary are close, and the nebula is large, and the CFW is
more than an order of magnitude faster than the slow wind, these
are very demanding simulations. 
\newline
{\bf Observations.}
The main issue is to detect binary companions.
Some methods with a table of 16 PNe with known central binary systems 
are given by Bond (2000, and references therein). 
Careful spectroscopic analysis can lead to more detection, as was done 
recently by Rodr\'{\i}guez et al.\ (2001).
X-ray observations can also lead to detection of companions.
An accreting  WD companion (Livio \& Shaviv 1975) will be too faint 
to be distinguish from the bright central stars of PNe.
However, spun-up main sequence stars can become magnetically active,
 and emit strongly in X-ray. This was suggested to be the case for the
point-like X-ray emission in the PNe NGC 6543 and NGC 7293 
(Guerrero et al.\ 2001).
Some magnetically active stars exist around WDs (e.g.,
Jeffries \& Stevens 1996; Bond et al.\ 2001).
The main sequence companions were spun-up by accreting from
the AGB progenitor, and then became magnetically active
(Jeffries \& Stevens 1996; Soker \& Kastner 2002). 
Therefore, I think that X-ray observations may reveal the existence of
main sequence companions to central stars of PNe
(Soker \& Kastner 2002).

{\bf Acknowledgments.}
I thank Guillermo Garc\'{\i}a-Segura for stimulating my research by
giving a good scientific fight back, with his high quality scientific work,
and by bringing different arguments for single star models during
our lively and friendly discussions.

\end{document}